\documentclass[sigconf]{acmart}
\AtBeginDocument{%
  }

\setcopyright{none} 
\copyrightyear{2025}
\acmYear{2025}
\acmConference[TREC iKAT'25]{TREC Interactive Knowledge Assistance Track}{August 2025}{Interactive and Offline Submission}

\usepackage{enumitem}
\usepackage{multicol}
\usepackage{multirow}
\usepackage{subcaption}
\usepackage{hyperref}

\usepackage{amssymb}
\usepackage{pifont}
\newcommand{\cmark}{\ding{51}}
\newcommand{\xmark}{\ding{55}}

\begin{document}

\title{CFDA \& CLIP at TREC iKAT 2025: Enhancing Personalized Conversational Search via Query Reformulation and Rank Fusion}
\renewcommand{\shorttitle}{Enhancing Personalized Conversational Search via Query Reformulation and Rank Fusion}
\author{Yu-Cheng Chang}
\affiliation{%
  \institution{CITI, Academia Sinica}
  \city{Taipei}
  \country{Taiwan}
}
\email{b12705055@ntu.edu.tw}
\authornote{These authors contributed equally to this work. Yu-Cheng Chang was a member of the interactive submission group and made the primary contribution to building the corresponding pipeline. Guan-Wei Yeo was a member of the offline submission group and made the primary contribution to building the pipeline. Both contributed to the writing of the final report.}

\author{Guan-Wei Yeo}
\affiliation{%
  \institution{CITI, Academia Sinica}
  \city{Taipei}
  \country{Taiwan}
}
\email{b11902091@csie.ntu.edu.tw}
\authornotemark[1]

\author{Quah Eugene}
\affiliation{%
  \institution{CITI, Academia Sinica}
  \city{Taipei}
  \country{Taiwan}
}
\email{216806@student.upm.edu.my}

\author{Fan-Jie Shih}
\affiliation{%
  \institution{CITI, Academia Sinica}
  \city{Taipei}
  \country{Taiwan}
}
\email{fanjie.shih@gmail.com}

\author{Yuan-Ching Kuo}
\affiliation{%
  \institution{CITI, Academia Sinica}
  \city{Taipei}
  \country{Taiwan}
}
\email{yckuo@citi.sinica.edu.tw}
\authornote{Leader of the interactive submission group; provided technical support and contributed to the writing of the final report.}

\author{Tsung-En Yu}
\affiliation{%
  \institution{CITI, Academia Sinica}
  \city{Taipei}
  \country{Taiwan}
}
\email{teyu@citi.sinica.edu.tw}
\authornote{Leader of the offline submission group; provided technical support.}

\author{Hung-Chun Hsu}
\affiliation{%
  \institution{CITI, Academia Sinica}
  \city{Taipei}
  \country{Taiwan}
}
\email{r10946017@citi.sinica.edu.tw}
\authornote{Overall leader and coordinator of the competition team; applied for datasets, ensured progress, guided technical direction, and contributed to revising the final report.}

\author{Ming-Feng Tsai}
\affiliation{%
  \institution{National Chengchi University}
  \city{Taipei}
  \country{Taiwan}
}
\email{mftsai@cs.nccu.edu.tw}

\author{Chuan-Ju Wang}
\affiliation{%
  \institution{CITI, Academia Sinica}
  \city{Taipei}
  \country{Taiwan}
}
\email{cjwang@citi.sinica.edu.tw}
\authornote{Corresponding author.}


\renewcommand{\shortauthors}{Chang et al.}

\begin{abstract}
The 2025 TREC Interactive Knowledge Assistance Track (iKAT) featured both interactive and offline submission tasks. The former requires systems to operate under real-time constraints, making robustness and efficiency as important as accuracy, while the latter enables controlled evaluation of passage ranking and response generation with pre-defined datasets. To address this, we explored query rewriting and retrieval fusion as core strategies. We built our pipelines around \textbf{Best-of-$N$ selection} and \textbf{Reciprocal Rank Fusion (RRF)} strategies to handle different submission tasks. Results show that reranking and fusion improve robustness while revealing trade-offs between effectiveness and efficiency across both tasks.
\end{abstract}





\maketitle
\section{Introduction}
Conversational search extends beyond traditional ad-hoc retrieval, where a retriever processes a single, well-formed query to retrieve relevant documents—by requiring systems to support dynamic, multi-turn interactions in which user intent evolves over time. In this setting, user queries are often ambiguous, incomplete, and context-dependent, necessitating systems to maintain dialogue state, infer implicit intent, and adapt to evolving user needs across turns. While prior research has relied on \textit{offline evaluation} \cite{aliannejadi_trec_2023, aliannejadi_trec_2024} with static datasets, this setup overlooks critical factors such as latency, robustness, and adaptation to unpredictable user behavior.

To address these limitations, the 2025 TREC iKAT introduced its first \textit{interactive submission task}, explicitly designed to motivate research communities to develop conversational retrieval systems capable of operating in realistic, real-time settings. In this task, a hidden large language model (LLM) simulates a human user through an API, producing dynamic and unpredictable interactions. Participants are tasked with building conversational retrieval systems that generate rewritten queries, retrieve relevant passages, and produce grounded responses in real time. These systems must also invoke and interact with the API at each step to accommodate evolving user inputs. This design aims to bridge the gap between controlled offline benchmarks and deployable conversational assistants by jointly evaluating both effectiveness and efficiency.

In this paper, we present our participation the 2025 TREC iKAT, covering both interactive and offline submission tasks. For the interactive submission, we submitted two runs: the first run combines CHIQ-AD \cite{mo-etal-2024-chiq} with LLM4CS \cite{mao-etal-2023-large} for adaptive query rewriting candidates using a \textbf{Best-of-$N$ selection} strategy, and the second run applies \textbf{Reciprocal Rank Fusion (RRF)} across multiple rewrites. For the offline submission, we submitted four automatic runs for passage ranking and response generation, each following a \textbf{Best-of-$N$ + RRF} strategy with slightly different configurations, as well as two generation-only runs. Our contributions are threefold: (i) we analyze how different rewriting strategies affect robustness under interactive evaluation, (ii) we demonstrate the effectiveness of retrieval fusion for improving coverage in both interactive and offline settings, and (iii) we discuss challenges unique to LLM-simulated evaluation, including latency optimization and context handling.

\section{Related Works}

Conversational search extends traditional ad-hoc retrieval, such as BM25 or sparse/dense vector retrieval, into multi-turn interactive scenarios where user queries are often ambiguous, incomplete, and context-dependent. Without considering dialogue history, retrieval systems frequently fail to capture the user’s true intent, leading to irrelevant or misleading results \cite{mao-etal-2023-large,mo-etal-2024-chiq,zhu_convsearch-r1_2025}. To address this challenge, two main paradigms have emerged: \textit{Conversational Dense Retrieval (CDR)} and \textit{Conversational Query Reformulation (CQR)}.

\subsection{Conversational Dense Retrieval}
Conversational Dense Retrieval (CDR) encodes both the dialogue history and the current user query into a dense representation using pre-trained encoders \cite{mo-etal-2024-history,zhu_convsearch-r1_2025,lai-etal-2025-adacqr}. By jointly modeling the query and its preceding turns, CDR is able to capture ellipsis and coreference within a conversation, enabling the retriever to produce context-aware relevance scores. 
This design allows CDR to be trained in an end-to-end manner, directly optimizing the encoder representations for conversational retrieval tasks.

Despite its effectiveness, CDR faces several limitations. 
First, these methods typically require costly fine-tuning of large transformer encoders on conversational datasets with dense relevance annotations, which restricts their applicability in domains where labeled data is scarce. 
Second, incorporating the full dialogue history often introduces redundant or noisy information, potentially degrading retrieval performance. 
To address this, several works propose techniques like history selection \cite{mo-etal-2024-history}, which reduce context length while preserving salient information. 
Nevertheless, the dependency on large-scale supervised datasets and task-specific fine-tuning remains a key bottleneck, limiting the scalability and generalization of CDR compared to lighter-weight alternatives such as conversational query reformulation (CQR).

\subsection{Conversational Query Reformulation (CQR)}
Unlike Conversational Dense Retrieval (CDR), which encodes the dialogue history and current query jointly, Conversational Query Reformulation (CQR) adopts a rewriting approach: the current user utterance is reformulated into a self-contained query that incorporates the necessary conversational context \cite{jang-etal-2024-itercqr,lai-etal-2025-adacqr,mao-etal-2023-large,mo-etal-2023-convgqr,mo-etal-2024-chiq,QRACDR}. 
By transforming the ambiguous or incomplete utterance into a stand-alone query, CQR enables the use of existing ad-hoc retrievers without requiring model re-training. 

CQR is generally more lightweight and practical than CDR, since it decouples rewriting from retrieval and can directly leverage strong off-the-shelf retrievers. 
However, the effectiveness of CQR depends heavily on the quality of the rewritten query. 
If the rewrite fails to capture subtle intent or misrepresents the conversation, important nuances may be lost, leading to retrieval errors. 
Recent work addresses these issues through large language models, reasoning-augmented reformulation, or history summarization \cite{jang-etal-2024-itercqr,mao-etal-2023-large,mo-etal-2024-chiq}, which have shown strong improvements over earlier rule-based or supervised rewriting methods. The following are several representative approaches illustrate the evolution of CQR: 

\subsubsection{LLM4CS Framework}
A widely adopted approach in CQR is the \textbf{Rewriting-And-Response (RAR)} prompting method, exemplified by the LLM4CS framework \cite{mao-etal-2023-large}. 
LLM4CS leverages large language models with chain-of-thought prompting to generate more accurate reformulations of conversational queries. 
This approach has been shown to better handle coreference and implicit intents compared to direct LLM rewrites, highlighting the potential of reasoning signals in query reformulation.

\subsubsection{CHIQ Framework}
The CHIQ framework \cite{mo-etal-2024-chiq} introduces a two-step process for conversational query rewriting. 
First, the dialogue history is enhanced through multiple LLM-based strategies (e.g., question disambiguation, response expansion, pseudo response, topic switch detection, and history summarization), which reduce ambiguity and noise. 
Based on this refined history, three variants are proposed for query rewriting.

\begin{itemize}[noitemsep, topsep=0pt, leftmargin=*]
    \item \textbf{CHIQ-AD} adopts a prompt-based rewriting strategy, directly leveraging enhanced history to generate reformulated queries. This reduces redundancy and helps prevent error accumulation in long dialogues.
    \item \textbf{CHIQ-FT} instead fine-tunes a lightweight T5-base model for rewriting. Its main advantage is efficiency during inference, but it suffers from a restricted context window (512 tokens), which limits its ability to capture long dialogue dependencies.
    \item \textbf{CHIQ-Fusion} combines the ranked lists retrieved from CHIQ-AD and CHIQ-FT using result-level fusion. This hybrid approach aims to integrate the robustness of prompt-based rewriting and the efficiency of fine-tuned models, showing stronger performance than either component alone in the original paper.
\end{itemize}

\subsection{Reward-based Query Reformulation}
Another line of research focuses on reward-based methods that align query reformulation with retrieval objectives. \textbf{AdaRewriter} \cite{lai_adarewriter_2025} adopts an inference-time strategy known as Best-of-$N$ decoding.
Given a conversational query, the system generates multiple candidate rewrites and uses a reward model—trained with a contrastive ranking loss on retrieval signals—to score them.  
The highest-scoring candidate is then selected as the final query.  
This design avoids the need for expensive fine-tuning and demonstrates the benefits of lightweight, reward-guided adaptation at test time.

In contrast, \textbf{ConvSearch-R1} \cite{zhu_convsearch-r1_2025} incorporates reinforcement learning to more tightly connect rewriting with downstream retrieval.  
It introduces a two-stage process: first, \textit{Self-Driven Policy Warm-Up} pre-trains the rewriter through self-distillation to mitigate cold-start issues; second, \textit{Retrieval-Guided RL} with rank-incentive reward shaping directly optimizes query rewrites using retriever feedback.  
This approach highlights the potential of reinforcement learning to align LLM-based rewriting with retrieval effectiveness more explicitly.

Together, AdaRewriter and ConvSearch-R1 illustrate the emerging trend of \textit{reward-based query reformulation}, where candidate generation is guided or optimized through retrieval-oriented signals, either at inference time or during training.

\subsection{Datasets and Benchmarks}
Progress in conversational information retrieval has been driven by several benchmark datasets.  
The earliest large-scale benchmark was \textbf{TREC CAsT (2019--2022)} \cite{dalton_trec_2020}, which introduced multi-turn dialogues for passage retrieval and established standard offline evaluation protocols.  
Subsequent datasets such as \textbf{QReCC} \cite{anantha-etal-2021-open} and \textbf{TopiOCQA} \cite{adlakha2022topiocqa} extended this setting by focusing on query rewriting and topic-oriented conversations, providing larger-scale training and evaluation resources.  
More recently, the \textbf{iKAT} track \cite{aliannejadi_trec_2023,aliannejadi_trec_2024} advanced the field by incorporating personalization through user personas and a Personal Text Knowledge Base (PTKB).  
While iKAT 2023–2024 remained offline in nature, the latest \textbf{iKAT 2025} introduced an \textit{interactive submission task}, where systems interact with an API in real time.  
This interactive task represents a significant shift in evaluation methodology, emphasizing latency, robustness, and efficiency trade-offs in addition to retrieval accuracy.
\section{Methodology}
\label{sec:method}

We model the iKAT 2025 submission tasks as a \textit{unified conversational retrieval pipeline}, applicable to both interactive and offline submissions. At each turn, given dialogue history $H$, current user utterance $u$, and a Personal Text Knowledge Base (PTKB) $K$, the system must generate a response grounded in retrieved evidence. Formally, the process can be expressed as:
\[
f_{\text{pipeline}} : (H, u, K) \mapsto (r, D, P),
\]
where $r$ is the generated response, $D$ is the ranked list of retrieved passages, and $P$ is the set of PTKB statements predicted as relevant. Our pipeline consists of four major stages: (i) query rewriting, (ii) passage retrieval, (iii) passage reranking, and (iv) response generation. These stages are described in detail below.  

\subsection{Query Rewriting}
\label{sec:candidate-generation}
The first stage of our pipeline reformulates the user's utterance $u$ conditioned on dialogue history $H$ and PTKB $K$ into a self-contained query $q'$. While both LLM4CS \cite{mao-etal-2023-large} and CHIQ \cite{mo-etal-2024-chiq} aim at generate contextually coherent rewrites, they differ in how conversational context is incorporated and how the rewriting process is optimized. In addition, AdaRewriter \cite{lai_adarewriter_2025} adaptively selects the most promising query from a set of candidate rewrites using an outcome-supervised reward model.

\subsubsection{LLM4CS Framework}
LLM4CS reformulates queries as:
\[
q' = f_{\text{RAR}}(H, u, K),
\]
where $f_{\text{RAR}}$ is a \textit{Rewriting-And-Response (RAR)} prompting process with a tailored Chain-of-Thought. This reformulation explicitly resolves ellipsis and coreference before generating a self-contained query. Compared with the original work, we omit the pseudo response (PR) component, as it often introduces noise and drifts the rewritten query away from the user’s true intent.

\subsubsection{CHIQ Framework}
In our work, we mainly adopt \textbf{CHIQ-AD} from the CHIQ framework. CHIQ-AD reformulates queries via a two-stage process:
\[
q' = f_{\text{rewrite}}\big(f_{\text{enhance}}(H, u), K\big),
\]
where $f_{\text{enhance}}$ performs history enhancement using five strategies: topic switch (TS), question disambiguation (QD), response expansion (RE), pseudo response (PR), and history summary (HS). The enhanced context is then passed to $f_{\text{rewrite}}$, which generates the final query. In our adaptation, we remove the history summary (HS) step, since it may drop important details in long conversations. Instead, we preserve the complete dialogue context.

\subsubsection{AdaRewriter Framework}
AdaRewriter introduces a reward-based inference-time adaptation. Rather than returning only a single LLM-produced rewrite, it selects the best query from a candidate set $\mathcal{Q} = \{q_1, q_2, \dots, q_N\}$ according to the reward model’s preference:
\[
q' = \arg\max_{q_i \in \mathcal{Q}} f_{\mathrm{reward}}(q_i, H, K),
\]
where $f_{\mathrm{reward}}$ is the score assigned by the reward model $M$. The query with the highest score is selected as the final query. Training details of the reward model are provided in Appendix~\ref{sec:train-adarewriter}.

\subsection{Passage Retrieval}
Given the rewritten query $q'$, the system retrieves an initial set of passages $D_0$ from the collection $\mathcal{C}$. Formally,
\[
D_0 = f_{\text{retrieve}}(q', \mathcal{C}),
\]
where $f_{\text{retrieve}}$ can be implemented with either sparse or dense retrieval methods.  
In our work, we adopt SPLADE \cite{splade}, a sparse expansion model shown to be effective in conversational and ad-hoc retrieval.  
Similar retrievers have been widely used in prior works \cite{pyserini}.

\subsection{Passage Reranking}
The initial set $D_0$ is further refined by a neural reranker $f_{\text{rerank}}$ to obtain the final ranking $D$. Formally,
\[
D = f_{\text{rerank}}(q', D_0).
\]
We consider multiple cross-encoder rerankers, including DeBERTaV3 and BGE, which differ in backbone architecture and training data.  
Rerankers consistently improve retrieval precision by modeling fine-grained interactions between $q'$ and candidate passages.  

\subsection{Response Generation}
We consider two approaches for response generation, corresponding to the interactive task and the offline task.

\subsubsection{Interactive Task}
The ranked list $D$ and the PTKB provenance $P$ (i.e., the set of relevant PTKB statements) are passed to an LLM $f_{\text{resp}}$, which generates the final response $r$:  
\[
r = f_{\text{resp}}(H, u, q', D_{1:10}, P),
\]
where $q'$ is the rewritten query and $D_{1:10}$ refers to the top- 10 passages from $D$. In practice, our implementation concatenates $q'$ and $D_{1:10}$ into the input prompt of LLM, together with $P$. This stage closes the loop of the interactive pipeline, as the generated $r$ is returned to the iKAT API, which subsequently provides the next user utterance.

\subsubsection{Offline Task}
In this task, response generation is divided into three stages: PTKB classification, passage summarization, and final response generation. At each turn, the PTKB is dynamically updated by detecting whether the current user utterance $u$ introduces new personal information not yet stored in $K_{t-1}$, yielding an updated $K_t$. A subset of relevant statements $P$ is then selected from $K_t$ to guide personalized response generation.

The reranked passage list $D$ is filtered by retaining the top-$\mathrm{NUM\_PASSAGES}$ passages whose relevance scores exceed the predefined $\mathrm{SCORE\_THRESHOLD}$. The top-$\mathrm{NUM\_DIRECT\_PASSAGES}$ passages are then preserved as direct inputs, while the remaining passages are grouped into chunks of size $\mathrm{SUMMARY\_CHUNK\_SIZE}$ and summarized by a large language model into a set of concise passages $D'$. Finally, these processed inputs are used to generate the final response. Formally,
\[
r = f_{\text{resp}}(H, u, D', P),
\]

\section{Experiments}
\label{sec:experiments}

\subsection{Datasets}
We evaluate our systems using both offline and interactive benchmarks. For offline evaluation, we leverage the iKAT 2023 and 2024 \cite{aliannejadi_trec_2023,aliannejadi_trec_2024} datasets, which provide multi-turn dialogues along with associated PTKB, enabling controlled analysis of query rewriting and passage retrieval methods. The same datasets are also used for interactive development and validation, allowing us to evaluate real-time system behavior in a simulated setting.

\subsection{Evaluation Metrics}
Following the iKAT guidelines, we report nDCG@10 and MRR@1000.  
nDCG@10 reflects the graded relevance of top-ranked passages, while MRR@1000 emphasizes early precision.
In the interactive task, additional aspects such as latency and robustness to dialogue variations are implicitly assessed.

\subsection{Baselines}
We compare our submissions against several baselines composed of various combinations of query rewriting, reranking, and fusion methods, detailed below:

\begin{itemize}[noitemsep, topsep=0pt, leftmargin=*]
    \item \textbf{Retriever-only:} A SPLADE retriever without any query rewriting or reranking, serving as the lowest-level baseline.  
    \item \textbf{Single-query rewriting:} Systems where a single rewriting
    \\strategy---GPT-4o mini, CHIQ-AD \cite{mo-etal-2024-chiq}, or LLM4CS \cite{mao-etal-2023-large} ---is used in combination with SPLADE \cite{splade} retrieval.  
    \item \textbf{Reranking:} Neural cross-encoders (DeBERTaV3 and BGE) applied on top of the retrieved passages to refine the initial ranking.  
    \item \textbf{Fusion:} Reciprocal Rank Fusion (RRF) over multiple rewrites, with two configurations: fusion applied \textit{before} reranking and fusion applied \textit{after} reranking.  
\end{itemize}

\subsection{Submission Runs}

\begin{figure*}[ht!]
  \centering
  \begin{subfigure}[t]{0.9\textwidth}
    \centering
    \includegraphics[width=\textwidth]{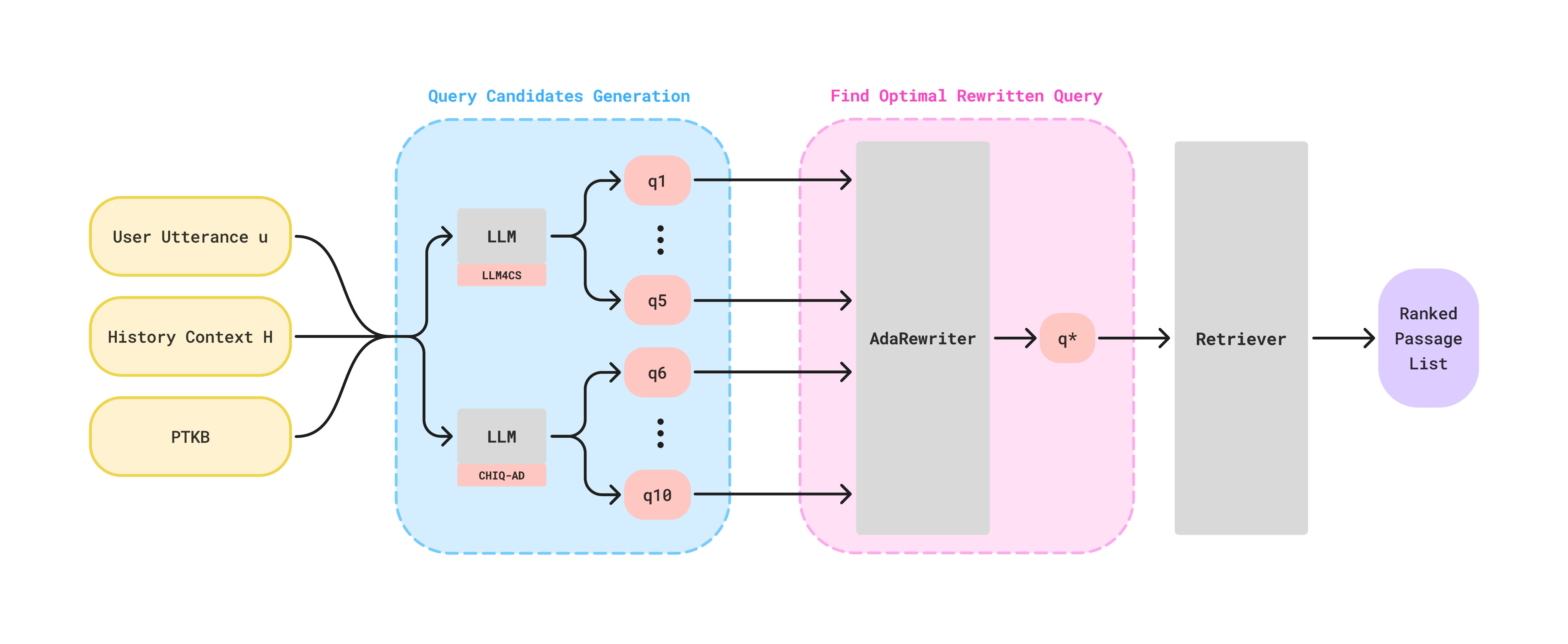}
    \caption{Interactive Run \#1: Best-of-$N$ Selection + Rerank pipeline.}
    \label{fig:run1}
  \end{subfigure}
  \vspace{1em}
  \begin{subfigure}[t]{0.9\textwidth}
    \centering
    \includegraphics[width=\textwidth]{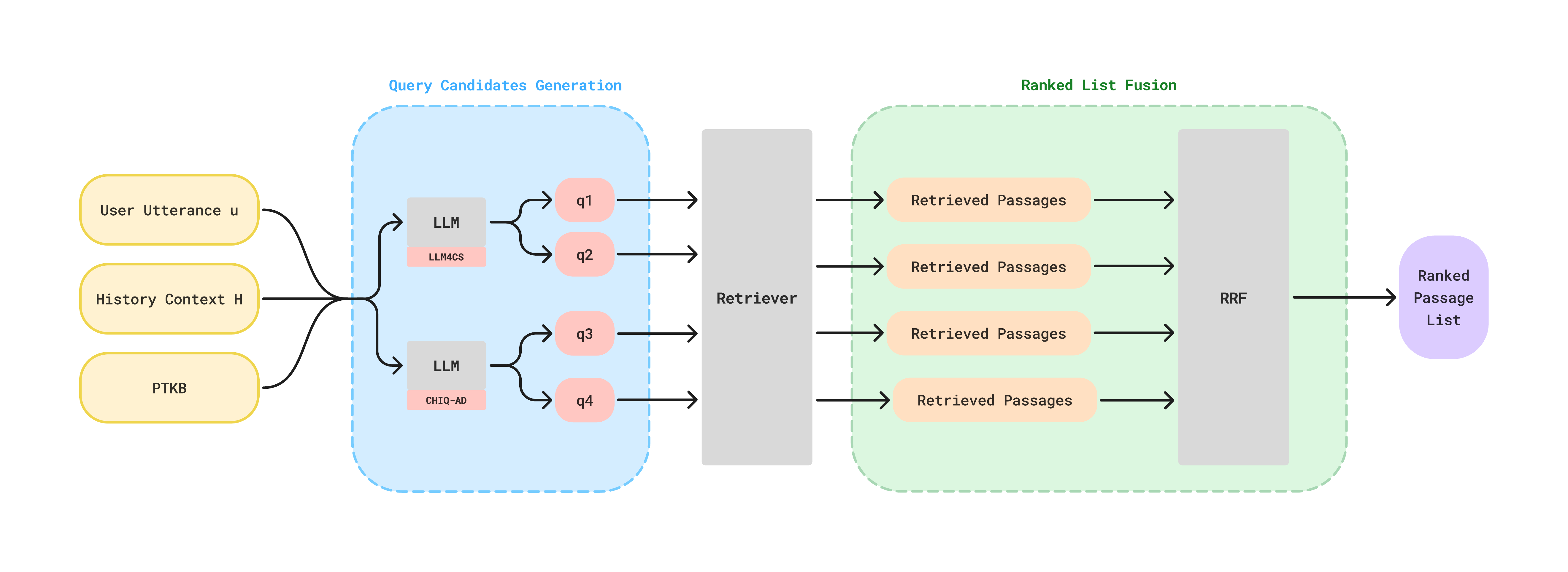}
    \caption{Interactive Run \#2: RRF + Rerank pipeline.}
    \label{fig:run2}
  \end{subfigure}
  \begin{subfigure}[t]{0.9\textwidth}
    \centering
    \includegraphics[width=\textwidth]{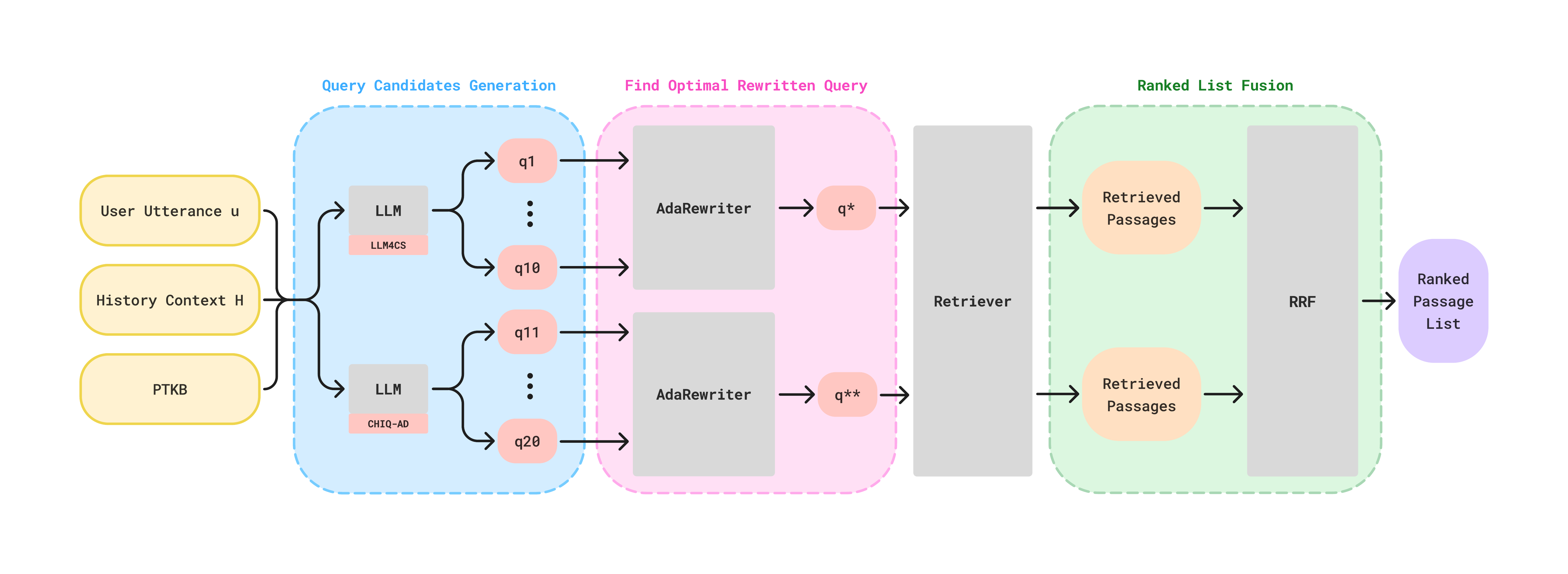}
    \caption{Offline All Runs: Best-of-$N$ Selection + RRF + Rerank pipeline.}
    \label{fig:run3}
  \end{subfigure}
  \caption{Submission pipelines for iKAT 2025. 
  (a) Interactive Run \#1: selects the best rewritten query via Best-of-$N$ Selection before retrieval and reranking. 
  (b) Interactive Run \#2: fuses passage rankings from multiple rewrites using RRF before reranking. 
  (c) Offline All Runs: integrates the query rewriting, passage retrieval, and reranking steps from both interactive pipelines.}
  \Description{Three subfigures illustrating the iKAT 2025 submission pipelines. 
  (a) Run 1: Best-of-$N$ Selection + Rerank for interactive evaluation. 
  (b) Run 2: RRF + Rerank for interactive evaluation. 
  (c) Offline runs: Best-of-$N$ Selection + RRF + Rerank for offline automatic evaluation.}
  \label{fig:pipelines}
\end{figure*}

For the iKAT 2025, we submitted runs for both the interactive and offline tasks: 2 interactive runs, 4 automatic runs, and 2 generation-only runs. The configuration of each run is described in detail below, while figure~\ref{fig:pipelines} summarizes the three distinct pipelines we adopted.

\subsubsection{Interactive Task Runs}
\leavevmode\\
\noindent \textbf{Run \#1: Best-of-$N$ Selection + Rerank.}
\label{par:online_run1}
This run integrates CHIQ-AD and LLM4CS for candidate query generation. Multiple rewrites are produced, and a pre-trained reward model that shares the same architecture as AdaRewriter's is employed to select the best candidate query. The selected query is then passed to SPLADE for top-2000 passage retrieval, followed by reranking with a cross-encoder (DeBERTaV3 or BGE). The final top-10 passages, together with the rewritten queries and PTKB provenance, are provided to the LLM for grounded response generation.

\noindent \textbf{Run \#2: RRF + Rerank.}
\label{par:online_run2}
This run employs parallel candidate generation with CHIQ-AD and LLM4CS, and each branch retrieves its own top-2000 passages with SPLADE. The two ranked lists are merged via Reciprocal Rank Fusion (RRF), then refined by a cross-encoder reranker (DeBERTaV3 or BGE). The final top-10 passages, along with the rewritten queries and PTKB provenance, are provided to the LLM for grounded response generation.

\subsubsection{Offline Automatic Task Runs}
\label{par:offline_run}
\leavevmode\\
\textbf{All Runs:  Best-of-$N$ Selection + RRF + Rerank.}
All four offline automatic runs share a unified pipeline. For each query, LLM4CS and CHIQ-AD each generate 10 candidate rewrites, and a pre-trained reward model that shares the same architecture as \\AdaRewriter's is employed to selects the best candidate from each set. The two selected queries are then used to retrieve the top-1000 passages via SPLADE. The resulting ranked lists are merged using Reciprocal Rank Fusion (RRF) and subsequently reranked with DeBERTaV3. For response generation, we adopt the default settings: $\mathrm{NUM\_PASSAGES}=20$, $\mathrm{SCORE\_THRESHOLD}=0.3$, $\mathrm{NUM\_DIRECT\_PASSAGES}=3$, $\mathrm{SUMMARY\_CHUNK\_SIZE}=5$.

The individual runs differ only in candidate configurations. \textbf{Run \#1:} CHIQ-AD without HS fused with the original LLM4CS. \textbf{Run \#2 and \#3:} CHIQ-AD without HS fused with LLM4CS without PR; the two runs use distinct candidate sets for robustness. \textbf{Run \#4:} Same as above, but with more diverse candidate variations.

\subsubsection{Offline Generation-Only Task Runs}
\leavevmode\\
Two generation-only runs differ only in passage filtering parameters. 
\textbf{Run \#1:} $\mathrm{NUM\_PASSAGES}=20$, $\mathrm{SCORE\_THRESHOLD}=0.3$, $\mathrm{NUM\_DIRECT\_PASSAGES}=3$, $\mathrm{SUMMARY\_CHUNK\_SIZE}=5$. 
\textbf{Run \#2:} $\mathrm{NUM\_PASSAGES}=13$, $\mathrm{SCORE\_THRESHOLD}=0$, $\mathrm{NUM\_DIRECT\_PASSAGES}=4$, $\mathrm{SUMMARY\_CHUNK\_SIZE}=5$. 

\section{Evaluation Results}
\label{sec:results}
\subsection{Main Results}
Table~\ref{tab:main-results} presents the performance of our submitted systems alongside key baselines on the \textbf{iKAT 2023} and \textbf{iKAT 2024}  \cite{aliannejadi_trec_2023,aliannejadi_trec_2024} datasets. Baselines include a retriever-only system (SPLADE \cite{splade}) and single-query rewriting methods (CHIQ-AD \cite{mo-etal-2024-chiq} and LLM4CS \cite{mao-etal-2023-large}). We report nDCG@10 and MRR@1K, highlighting the best scores in bold and the second-best scores with underlines. These results reveal three main observations:

\begin{table*}[ht]
\centering
\caption{Main results on the iKAT evaluation sets. We report nDCG@10 and MRR@1K. The table includes a retriever-only baseline, single-query rewriting baselines, and our final systems. Best results in \textbf{bold}, second-best underlined.}
\label{tab:main-results}
\begin{tabular}{lccccccccc}
\toprule
\multirow{2}{*}{\textbf{Method}} 
& \multicolumn{4}{c}{\textbf{Retrieval Pipeline Components}} 
& \multicolumn{2}{c}{\textbf{iKAT-2023}} 
& \multicolumn{2}{c}{\textbf{iKAT-2024}} \\
\cmidrule(lr){2-5} \cmidrule(lr){6-7} \cmidrule(lr){8-9}
 & \textbf{Rewrite} & \textbf{\#Queries} & \textbf{Reranker} & \textbf{RRF} 
 & \textbf{nDCG@10} & \textbf{MRR@1K} 
 & \textbf{nDCG@10} & \textbf{MRR@1K} \\
\midrule
SPLADE \cite{splade} & \xmark & Single & \xmark & \xmark & 0.1849 & 0.2867 & 0.2974 & 0.4881 \\
CHIQ-AD \cite{mo-etal-2024-chiq} & \cmark & Single & \xmark & \xmark & 0.1753 & 0.2638 & 0.3221 & 0.5422 \\
LLM4CS \cite{mao-etal-2023-large} & \cmark & Single & \xmark & \xmark & 0.2118 & \textbf{0.3463} & 0.3230 & 0.5036 \\
\midrule
Interactive Run\#\hyperref[par:online_run1]{1} (ours) & \cmark & Multiple & \cmark & \xmark & 0.2153 & 0.3134 & 0.4218 & \underline{0.6646} \\
Interactive Run\#\hyperref[par:online_run2]{2} (ours) & \cmark & Multiple & \cmark & \cmark & \textbf{0.2215} & \underline{0.3415} & \textbf{0.4425} & 0.6629 \\
Offline All \hyperref[par:offline_run]{Runs} (ours) & \cmark & Multiple & \cmark & \cmark & \underline{0.2161} & 0.3174 & \underline{0.4303} & \textbf{0.6721} \\
\bottomrule
\end{tabular}
\end{table*}

\begin{itemize}[noitemsep, topsep=0pt, leftmargin=*]
    \item Both CHIQ-AD and LLM4CS outperform the retriever-only baseline, confirming the importance of context-aware query rewriting for conversational search.  
    \item RRF-based fusion achieves the strongest overall performance, demonstrating that combining multiple rewrites improves robustness to conversational ambiguity.
    \item While LLM4CS yields the highest single-query MRR@1K on iKAT 2023, our fusion system delivers more consistent performance across datasets, validating the robustness of the fused approach.  
\end{itemize}

\begin{table}[t!]
\centering
\caption{Baseline results on iKAT 2024 with single-query rewriting.}
\label{tab:ikat2024-baseline}
\begin{tabular}{l l cc}
\toprule
\textbf{Model} & \textbf{Rewrite} & \textbf{nDCG@10}  & \textbf{MRR@1K} \\
\midrule
SPLADE & GPT-4o mini & 0.2974 & 0.4881 \\
\quad+ DeBERTaV3 & GPT-4o mini & 0.3923  & 0.6128 \\
\midrule
SPLADE & CHIQ-AD & 0.3221  & 0.5422 \\
\quad+ DeBERTaV3 & CHIQ-AD & \underline{0.4112}  & \underline{0.6334} \\
\midrule
SPLADE&LLM4CS & 0.3230  & 0.5036 \\
\quad+ DeBERTaV3 & LLM4CS & \textbf{0.4432}  & \textbf{0.6396} \\
\midrule
SPLADE & CHIQ-FT\cite{mo-etal-2024-chiq} & 0.1286 & 0.2304\\ 
\bottomrule
\end{tabular}
\end{table}

\begin{table}[t!]
\centering
\caption{RRF results on iKAT 2024 combining LLM4CS and CHIQ-AD rewrites under different reranking configurations.}
\label{tab:ikat2024-rrf}
\begin{tabular}{l l cc}
\toprule
\textbf{Model} & \textbf{Config} & \textbf{nDCG@10} &  \textbf{MRR@1K} \\
\midrule
SPLADE & RRF only & 0.2227  & 0.3337 \\
\quad+BGE & RRF first & 0.4115  & 0.6135 \\
\quad+DeBERTaV3 & RRF first & \textbf{0.4425}  & \textbf{0.6629} \\
\quad+DeBERTaV3 & Rerank first & \underline{0.4275}  & \underline{0.6476} \\
\bottomrule
\end{tabular}
\end{table}

\subsection{Ablation Studies}
To systematically isolate and evaluate the contributions of individual components, we conducted a series of ablation studies on the iKAT 2024 dataset. Specifically, we investigate: (i) the effect of query rewriting strategies, (ii) the impact of neural rerankers, (iii) the ordering of RRF fusion and reranking, (iv) the limitations of fine-tuned alternatives such as CHIQ-FT, and (v) the influence of the number of candidates $N\_{\text{candidates}}$ in AdaRewriter. Experimental results are summarized in Tables~\ref{tab:ikat2024-baseline} and~\ref{tab:ikat2024-rrf}, as well as Figure~\ref{fig:ikat2024-n_cand}, from which several key observations can be drawn.

\subsubsection{Effect of Query Rewriting}  
Table~\ref{tab:ikat2024-baseline} shows that advanced query rewriting consistently improves retrieval performance over simple prompting.  
Compared to the lightweight GPT-4o-mini baseline, both CHIQ-AD and LLM4CS achieve higher scores in nDCG@10 and MRR@1K.  
This confirms that leveraging dialogue history (CHIQ-AD) or reasoning-augmented reformulation (LLM4CS) is more effective than simply prompting a LLM to obtain  queries that can better capture user intent in conversational search scenarios. 

\subsubsection{Impact of Reranking}  
Table~\ref{tab:ikat2024-baseline} also shows that adding a reranker consistently improves performance across all rewriting strategies.  
For example, when using rewritten queries from GPT-4o-mini, the reranker DeBERTaV3 increases nDCG@10 scores from 0.2975 to 0.3923, while with rewritten queries from CHIQ-AD, DeBERTaV3 improves scores from 0.3221 to 0.4112.
This highlights the crucial role of neural rerankers in improving retrieval precision.

Table~\ref{tab:ikat2024-rrf} further compares different reranker backbones under RRF fusion. While DeBERTaV3 achieves the highest scores, BGE performs slightly lower but remains competitive. These results indicate that, although absolute effectiveness depends on the backbone, reranking consistently improves performance when applied after query fusion.

\subsubsection{Ordering of RRF and Reranking}  
Table~\ref{tab:ikat2024-rrf} shows that applying RRF before reranking outperforms the reverse order. This indicates that fusing rewrites prior to reranking exposes the reranker to a richer and more diverse candidate set, improving robustness to query ambiguity and top-$k$ ranking quality.

\subsubsection{Limitations of CHIQ-FT}  
As shown in the last row of Table~\ref{tab:ikat2024-baseline}, we also evaluated the effectiveness of \textbf{CHIQ-FT}, which generates queries using a small LM (e.g., T5-base) fine-tuned for conversational query reformulation.  
Its performance was noticeably worse than prompt-based rewriting.  
The main limitation lies in its restricted input length (i.e., 512 tokens), which often truncates long dialogue histories and leads to incomplete or inaccurate rewrites.  
This illustrates a broader trade-off between efficiency and robustness in lightweight fine-tuned models.

\subsubsection{Effect of the number of candidates in AdaRewriter}  
\begin{figure}[t!]
\centering
\includegraphics[width=0.8\linewidth]{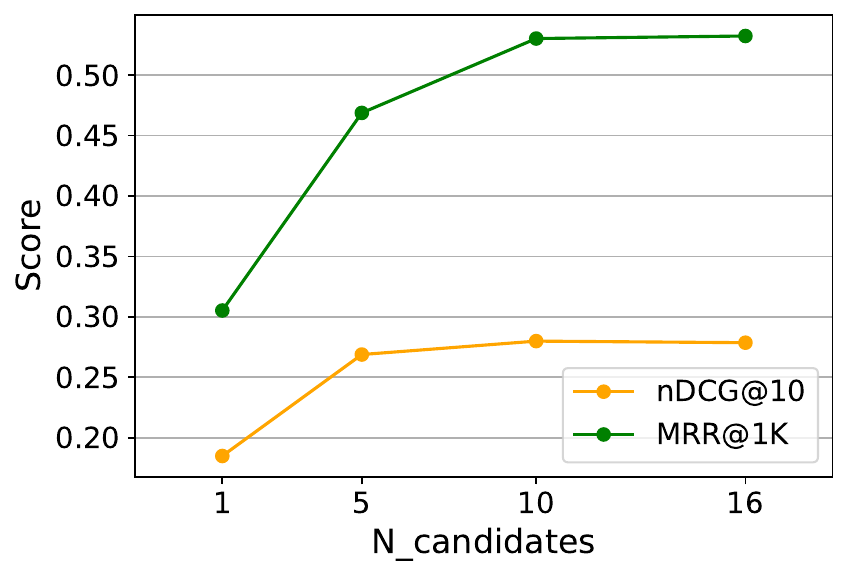}
\caption{Results on iKAT 2024 varying $N\_{\text{candidates}}$ in AdaRewriter. We report the average across all queries.}
\Description{Results on iKAT 2024 varying $N\_{\text{candidates}}$ in AdaRewriter. We report the average across all queries.}
\label{fig:ikat2024-n_cand}
\end{figure}

Figure~\ref{fig:ikat2024-n_cand} shows that retrieval performance generally increases as $N\_{\text{candidates}}$ grows. However, beyond $N\_{\text{candidates}}=10$, improvements are marginal while computational cost rises.  
Considering the trade-off between effectiveness and efficiency, we set $N\_{\text{candidates}}=10$ in all offline runs.

\subsubsection{Summary}  
Our experiments show that:    
(i) advanced query rewriting (CHIQ-AD, LLM4CS) significantly outperforms simple prompting;  
(ii) reranking is essential for boosting retrieval precision; and  
(iii) RRF fusion combined with reranking achieves the strongest overall performance.
These findings explain the effectiveness of our submitted runs: \textbf{Best-of-$N$ Selection + Rerank} leverages adaptive best-of-$N$ rewriting, while \textbf{RRF+Rerank} exploits fusion and reranking to maximize robustness under conversational ambiguity.

\section{Conclusion and Future Work}
In this work, we presented CFDA’s submission to the 2025 TREC iKAT 2025, exploring query rewriting and retrieval fusion for conversational search. Our experiments with AdaRewriter \cite{lai_adarewriter_2025}, CHIQ-AD \cite{mo-etal-2024-chiq}, and LLM4CS \cite{mao-etal-2023-large} demonstrate that carefully adapting rewriting strategies and fusion mechanisms can yield consistent improvements across evaluation sets.
Looking ahead, we identify several promising directions for future work.

\begin{enumerate}
    \item \textbf{Latency optimization.} Interactive evaluation imposes strict timing constraints; future systems should incorporate lightweight rerankers or early-exit mechanisms to balance effectiveness with response speed.
    \item \textbf{Adaptive query strategies.} Instead of always performing multi-query fusion, the system could dynamically decide between single-query retrieval and multi-query RRF based on query complexity or dialogue context.
    \item \textbf{Deeper PTKB integration.} Current usage of the PTKB is limited. Future work could explore entity linking and knowledge-grounded rewriting to more effectively incorporate PTKB entries into both query rewriting and response generation.
    \item \textbf{Enhanced passage summarization.} The response generation stage could benefit from more specialized summarization models or methods, such as PEGASUS \cite{10.5555/3524938.3525989}, BART \cite{lewis-etal-2020-bart}, LongT5 \cite{guo-etal-2022-longt5}, or extractive-abstractive hybrid pipelines tailored for long-form and domain-specific passages.
\end{enumerate}

In conclusion, our study underscores the importance of modular pipelines that balance rewriting, reranking, and fusion under interactive constraints. We hope these findings and future directions will inform the development of more adaptive and deployable conversational search systems.

\bibliographystyle{ACM-Reference-Format}
\bibliography{ref}

\appendix

\section{Dataset Details}
Table~\ref{tab:dataset-stats} summarizes the statistics of the three datasets used in our experiments: TREC iKAT 2023-2024 \cite{aliannejadi_trec_2023,aliannejadi_trec_2024}, and QReCC \cite{anantha-etal-2021-open}.

The iKAT datasets consist of multi-turn conversational queries paired with large-scale passage collections derived from ClueWeb22 \cite{10.1145/3477495.3536321}. We use these datasets exclusively for evaluation, as they allow us to assess the effectiveness of our proposed pipeline design.

In contrast, the QReCC dataset is used to train the reward model in the AdaRewriter module. It consists of 14K conversations with 81K question–answer pairs, paired with a collection of 54M passages. 
Due to its large scale, we rely on the processed passage set released by \cite{scai-qrecc} for practical training. Furthermore, since the original QReCC training set does not contain gold passage annotations, we adopt the supplementary labels provided by \cite{infocqr}.

\begin{table}[ht]
\centering
\caption{Statistics of TREC iKAT 2023-2024, and QReCC.}
\label{tab:dataset-stats}
\begin{tabular}{lccc}
\toprule
 & \textbf{iKAT-23} & \textbf{iKAT-24} & \textbf{QReCC (Train)} \\
\midrule
\# Dialogues & 13 & 17 & 10823 \\
\# Turns & 326 & 218 & 29596 \\
\midrule
\# Collections & 116M & 116M & 54M \\
\bottomrule
\end{tabular}
\end{table}

\section{Implementation Details}

\subsection{Query Rewriting}
We evaluate four query rewriting approaches.  
During development and ablation studies, all GPT-based methods are run with \textbf{GPT-4o-mini} for efficiency.  
For the official submissions, we employ the full \textbf{GPT-4} models in the online track.

\begin{itemize}
    \item \textbf{GPT-4o mini baseline:} Prompt-based rewriting using GPT-4o-mini with a fixed system prompt for conversational search.
    \item \textbf{CHIQ-AD:} Adaptive disambiguation strategy implemented with GPT-4 (GPT-4o-mini in ablations), following the CHIQ benchmark setup.
    \item \textbf{LLM4CS:} A GPT-4 based rewriting method (GPT-4o-mini in ablations) designed for large-scale conversational search, focusing on reasoning-augmented reformulation.
    \item \textbf{CHIQ-FT:} A fine-tuned T5-base model (non-GPT), included for comparison. Its shorter context length (512 tokens) limits its effectiveness in handling long dialogue histories.
\end{itemize}

\subsection{Training Details of the Reward Model}
\label{sec:train-adarewriter}
All experiments were conducted on a server equipped with NVIDIA V100 GPUs.  

We used the QReCC \cite{anantha-etal-2021-open} dataset and employed the vLLM \cite{vllm} framework to serve \texttt{LLaMA-3.1-8B-Instruct} for candidate generation, with $N\_{\text{candidates}}=5$ and temperature set to 1.0.  

Passage retrieval was performed using BM25 with parameters $(k_1, b)=(0.82, 0.68)$ via Pyserini \cite{pyserini} indexing. Here, $k_1$ controls non-linear term frequency normalization, while $b$ adjusts the scaling of the inverse document frequency. In our implementation, we remove the dense retrieval score term from the scoring function for ranking assessment, as passage retrieval in our pipeline relies solely on sparse scores.  

The AdaRewriter \cite{lai_adarewriter_2025} module was trained using PyTorch Lightning with the AdamW optimizer, a learning rate of 5e-6, and a cosine learning rate schedule with a warmup ratio of 0.1. Training was carried out for 13 epochs, with model checkpoints saved at the end of final epoch.

\end{document}